# Mitigating collusion in Gold Humanism Honor Society nominations

Congzhou M Sha, PhD[1,*]

[1]Medical Scientist Training Program, Penn State College of Medicine, Hershey, PA 17033

*Corresponding author. Email: cms6712@psu.edu, 500 University Dr, Hershey, PA 17033

## Abstract

An anecdotally common complaint regarding induction into the Gold Humanism Honor Society is the bias toward close friends during the initial nomination process. In this work, we numerically simulate the nomination process under different assumptions, demonstrate that collusion can be detected, and propose a simple strategy to correct for bias in the nomination process.

## Introduction

Common complaints regarding induction to the Gold Humanism Honor Society (GHHS) involve fairness[1], for example for MD-PhD students who may have smaller friend groups during their clinical years[2], or applicants to competitive specialties who may take a research year prior to their final year of medical school. To understand these complaints, we first note that specific selection criteria are suggested by the Arnold P. Gold Foundation, however it is up to the individual schools to decide how to implement these criteria. We will focus on the simplest part of the process for which there is a standardized method of assessment: the initial nomination of medical students by their peers.

The Gold Foundation recommends that medical students nominate up to eighteen distinct other students who come to mind, based on the six-item McCormack survey[3]. The responses are intended to be spontaneous, and campaigning is typically prohibited. Even in the absence of overtly dishonest behavior, certain biases can appear such as through friend group in-voting. In this work, we examine nomination process from the perspective of game theory, create a simple probabilistic model of voting, and propose a method of reducing the effect of collusion and campaigning with favorable theoretical and numerical results.

**Methods**

The Nash equilibrium in a simple case

Suppose we have a set of students who are all equally qualified for GHHS. What is the optimal strategy for a given student, if students are not allowed to collude (i.e. there are no friend groups)? Since students are not allowed to self-nominate, there is no incentive for a given student to nominate any of their peers since doing so would strictly worsen their own position in terms of vote counts. It is easy to see that the Nash equilibrium occurs when no student nominates any other student, and therefore all students have an equal number of votes (i.e. zero), with equal chances to be inducted.

Friend group collusion

Now we assume that collusion is present, and that there is no dishonesty within each friend group. Since there is a quota $q$ for the number of students allowed to be inducted, members of friend groups of up to size $q + 1$ may strategize to nominate all other members within that group, resulting in at least $q$ votes for each member, with additional votes coming from individuals outside of that friend group. If we now assume there are two friend groups $A$ and $B$ with sizes $a$ and $b$ respectively, and $a < b \leq q + 1$ which follow this strategy, then all members of group $B$ will have strictly more votes than all members of group $A$, resulting in group $B$ being ranked more highly than group $A$, despite equal qualifications.

This analysis motivates a possible requirement for a fair scoring system:

*The score should be independent of friend group size.*

This point can be argued if one believes that having a larger friend group automatically means that the student is more qualified, however this is a dangerous line of thought because it further isolates underrepresented students who do not have enough similar peers to form the robust friend groups described here.

<u>Detecting collusion</u>

To detect the collusion: suppose that there are $N$ students, that all students vote at random according to their true opinions on the humanistic quality of other students, and that all students are of equal humanistic quality. We expect that all $N$ students will have a roughly equal number of votes. What is the probability of the event $X$, that the same set of $n \leq 18$ students will be voted for by two other students, assuming both students nominated 18 people?

Assume that student A has already made their selection. The total number of possible choices for student B is $\binom{N}{18} = \frac{N!}{18!(N-18)!}$. The number of ways that student B chooses $n$ students that student A picked is $\binom{18}{n}$, implying that the remaining $18 - n$ students student B chooses are from the $N - 18$ students which student A did not choose, for which there are $\binom{N-18}{18-n}$ possibilities. Thus the probability that exactly $n$ students are in common is the familiar hypergeometric distribution:

$$P(X = n) = \frac{\binom{18}{n}\binom{N-18}{18-n}}{\binom{N}{18}}. \tag{1}$$

Therefore the probability that at least $n$ students are in common is:

$$P(X \geq n) = \sum_{k=n}^{18} \frac{\binom{18}{k}\binom{N-18}{18-k}}{\binom{N}{18}}. \tag{2}$$

We may now set a cutoff $\alpha$ such that $P(X \geq n) < \alpha$ implies that there was possible strategizing in voting involved, and thus appropriately identifying voting sheets for which the same $n$ candidates were voted for. We did not implement this method in our simulations because not all students were assumed to be equally qualified.

A probabilistic model of the GHHS nomination process

The preceding analyses were highly simplified models of the voting process. Here, we added some complexity to the voting process using a probabilistic model. We examine the limiting cases of the model analytically, and simulate it numerically.

For $N$ students in the medical school class, per the GHHS nomination rules, only the top $15\% \leq p \leq 17\%$ of students in the class may ultimately *be* inducted. Next, we assume that a fraction $f_{\text{deserving}} \geq p$ of the students are *deserving* of induction. Now we make the major simplifying assumption that the students who deserve induction are $r$ times as "humanistic" as those who are not, and we call $r_{\text{humanism}}$ the *maximum humanism ratio*, so in practice we assign the deserving students a *humanism score* of $r_{\text{humanism}}$ and all other students a humanism score of one.

We model the student body as a graph[4] $G = (V, E)$, where vertices represent students and edges represent friendship ties (i.e. possible collusion). We assign friend groups which form $n_c$ strongly-connected cliques, with the $i^{th}$ clique having size $c_i$, and with cliques otherwise being disjoint. Finally, to model outside-of-clique friendships, we assign friendships to a fraction $f_{\text{friendship}}$ of all other possible pairs of students. The *friend group* of a student $i$ is the set of neighbors of $i$ in the graph $G$.

Next, we describe the voting process. We assume that for each entry in the nomination form of 18 entries, students have a probability $f_{\text{clique}}$ of choosing a friend from their friend group (if not all friends have yet been nominated). We call $f_{\text{clique}}$ the *ingroup preference*. If a friend is not chosen for that entry, we assume that the student picks someone from outside of the friend group, with each student weighted according to their humanism score.

In a naïve scoring situation, the $\lfloor p \cdot N \rfloor$ students with the most votes (by counting the number of ballots which include those students) are inducted.

*Limiting case 1*

We analyze the case of perfect collusion ($f_{\text{clique}} = 1$) and clearly deserving students ($r_{\text{humanism}} \to \infty$). The most fair result is that each deserving student has a probability $\min\left(1, \frac{f_{deserving}}{p}\right)$ of being inducted, i.e. the deserving students are all chosen for nomination, or as many of them as possible are chosen where the probability of choosing each student is uniformly distributed. If there is perfect collusion and at least one non-deserving student is a member of the largest friend group, then the final tally deviates from the most fair result.

*Limiting case 2*

We analyze the case of no friend groups and finite humanism ratio $r_{\text{humanism}}$. Each student then chooses 18 others. In other words, we perform weighted drawing without replacement, which is described by Wallenius' noncentral hypergeometric distribution [5], whose probability mass function (PMF) is known [6] to be

$$\text{PMF}(x) = \binom{m_1}{x}\binom{m_2}{n-x} \int_0^1 \left(1 - t^{\frac{r_{\text{humanism}}}{D}}\right)^x \left(1 - t^{\frac{1}{D}}\right)^{n-x} dt, \qquad (3)$$

with $n = \lfloor pN \rfloor, m_1 = \lfloor f_{\text{deserving}} N \rfloor, m_2 = n - m_1, D = r_{\text{humanism}}(m_1 - x) + (m_2 - (n-x))$, and $x$ is the number of deserving students chosen. The $x$ votes are then distributed among the $m_1$ students uniformly. The mean $\bar{x}$ and variance $\sigma$ can be calculated directly through enumeration of the PMF. The case of multiple subpopulations with varying $r_{\text{humanism}}$ corresponds to the multivariate Wallenius' noncentral hypergeometric distribution.

Simulation of cliques

We assumed that the friend group sizes were repeatedly drawn from a normal distribution with mean of 10 and standard deviation of 6 (truncated at 0) until the sum of friend group sizes was greater than or equal to the number of students. We performed this drawing of friend groups 100 times, to simulate a range of friend group distributions, and took a median across these simulations. We also assumed that the deserving students were uniformly distributed across all students.

Collusion mitigation strategies

We examined two methods of collusion mitigation.

*Method 1: connected components on undirected graphs*

One method was to detect the connected components of the graph of votes[4], where an undirected edge $(a, b)$ represents that student $a$ voted for student $b$, and to penalize all votes in that connected component by the size of the connected component.

*Method 2: shared edges on directed graphs*

Another method was to consider the graph of votes as directed, discount all votes if both $(a, b)$ and $(b, a)$, i.e. students both voted for each other. We refer to this method as reciprocal vote discounting.

Software and statistical analysis

We used Julia 1.11.5 for all our code, which is included as Supplemental Material. For each set of parameters, one hundred simulations were performed. We used the two-tailed Mann-

Whitney U-test[7] for statistical significance, and set an overall significance level at $\alpha < 0.05$. To reduce false positives, we used a Bonferroni correction, resulting in $p < 0.000397$ as the level of significance.

**Results**

We may now examine the percentage of chosen students who are part of the deserving group. For simplicity, we chose the percentage of the class inducted as $p = 15\%$ for all the simulations described here. As our endpoint for fairness, we examined the fraction of chosen students $f_{\text{chosen \& deserving}}$ who were part of the $f_{\text{deserving}}$ fraction, i.e. deserving of the award. Higher values of $f_{\text{chosen \& deserving}}$ indicate that the voting and tallying process were better able to choose the truly outstanding students in the class. We closely examined three independent variables: the total number of students $N$ (50, 150, or 150), the fraction of the class which was deserving $f_{\text{deserving}}$ (0.15 to 0.40 in increments of 0.05), and the factor by which deserving students outperformed the rest $r_{\text{humanism}}$ (1.5, 2.0, 2.5, 3.0, 5.0, 7.5, or 10.0). We show the median $f_{\text{chosen \& deserving}}$ in the form of heatmaps which share single color scale (Figure 1, Figure 2, Figure 3, Figure 4).

First, we examined the case with no preference for friends, with no cliques and $f_{\text{clique}} = 0$ (Figure 1). In other words, students were nominated based on solely their humanistic quality. Reassuringly, the nomination process was highly sensitive and specific to the deserving nature of students, even when $r_{\text{humanism}}$ was low (i.e. the deserving students did not outperform other students by a large margin) and small class size.

Second, we tested simple tallying in the perfect collusion scenario with $f_{\text{clique}} = 1$ (Figure 2). Collusion introduced significant noise into the tallying process, resulting in far fewer chosen students being selected from the deserving population. This effect was somewhat mitigated at larger class sizes ($N$), for students who were highly deserving ($r_{\text{humanism}}$), and for a larger deserving population ($f_{\text{deserving}}$). We used this scenario as the control to test our two collusion mitigation strategies.

Third, we tested the connected component method, in which votes associated with connected components in the graph of votes were removed (Figure 3). This method did improve the fraction of deserving students $f_{\text{deserving}}$ who were chosen, but only when this fraction was sufficiently high. Interestingly, it decreased the fraction of deserving students chosen when $f_{\text{deserving}}$ was low.

Finally, we tested reciprocal vote discounting (Figure 4, Figure 5), indicating the statistically significant scenarios in Figure 6. This method yielded the best results, increasing the fraction of students chosen who were also deserving across the entire range of scenarios with few exceptions.

**Discussion**

There is a dearth of literature on the fairness of the GHHS nomination process. This work aims to provide theoretical justification that bias due to collusion is both impactful and can be easily detected and fixed. In a perfect world, students would vote according to their true opinions of their peers, however, medicine is a highly competitive environment which may encourage dishonest behavior[8,9]. If there is an opportunity to limit the impact of bad actors, we should take it.

In the Methods, we showed that the GHHS nomination process is a zero-sum game, since a student voting for others directly strictly decreases the probability that the student will be inducted ("The Nash equilibrium in a simple case"). We then provided a simple example in which friend group size biases the results of the nomination process ("Friend group collusion"). In the same setting, we showed how one might detect collusion using probabilities computed from the hypergeometric distribution ("Detecting collusion"), with the caveat that the underlying assumptions may not hold in more realistic settings.

We moved on to a probabilistic model which balanced simplicity with realism ("A probabilistic model of the GHHS nomination process"), conducting both theoretical and numerical analyses which are the focus of this study. The theoretical analysis took us as far as the distribution of votes made by each student across deserving and non-deserving students, in the absence of friend groups. We will leave the theoretical discussion at: the subsequent tallying and ranking process is difficult to study analytically.

In numerical simulations, we demonstrated the profound the effect that friend group collusion can have on tallies (Figure 1, Figure 2). One graph theoretical method to reduce votes was unsuccessful (Figure 3), however simply searching for reciprocal voting resulted in a robust method of significantly reducing the effect of collusion, both qualitatively (increasing the quality of inducted students in up to 30% of the chapter, Figure 5) and statistically (Figure 6).

Reciprocal voting is found when students campaign for themselves (which is expressly prohibited by the GHHS nomination rules) or when they vote for their friends. In the friend group setting, it is implicit that all friends will vote for all other friends, resulting in an increase in reciprocal voting. When the friend group is small, accordingly fewer reciprocal votes will be discounted. Similarly, when the class size is large, it is less likely for a given student to simultaneously be voted for by one of their nominees, resulting in robustness against collusion in the large class size limit. We contend that since reciprocal voting detection can in principle be performed via pencil and paper, and due to the encouraging numerical results presented herein, it is an appropriate for implementation at schools with a GHHS chapter.

One may argue that further screening measures in the nomination process such as interviews may highlight deserving students, however in this case there would be a significant fraction of deserving students who were unfairly screened out at an earlier stage. The solution is either to increase the number of students interviewed, which may be burdensome on the chapter, or to use some means to ensure specificity, such as reciprocal voting detection.

The presence of friend groups in each medical school class can only be ascertained by direct survey of students. However, we believe it would be unreasonable to say that friend groups do not exist and do not influence voting patterns. To give an example in which collusion is deeply troubling, consider the case of ethnic minorities. Shunning from large friend groups would dramatically increase the difficulty with which such students would be awarded GHHS membership. Since GHHS status has a nonzero effect on residency applications [1], such bias may contribute to systemic discrimination against ethnic minorities.

Regarding limitations of this study, we first note that all computational studies including this one are speculative and depend on a variety of assumptions. Here, we assumed that there was such a thing as an objective measure of humanistic quality ($r_{\text{humanism}}$), however we would argue that this is an assumption of the GHHS as well, as it precisely seeks to distinguish these students. One can certainly assign scores of varying objectivity to humanistic quality, e.g. number of hours volunteered, number of people in the community reached, or interview scores. We assumed a simple stratification of the class into deserving students and other students, and that students will always prioritize their friend groups over others. Additionally, we assumed that every student cast all 18 of their votes. Future studies on this subject may consider introducing additional complexity into voting simulation, or correlating real-world data with simulation parameters to validate the methods outlined here.

The reciprocal voting method proposed is still prone to dishonest behavior, but it significantly increases the overall effort of that behavior. To collude on votes in a fair market, a student would have to ensure that votes are not traded directly and is traded instead through a third-party proxy. In principle, a centralized vote trader could circumvent reciprocal voting detection. However, we note that realistically, many individuals would need to be involved, presumably significantly increasing the risk of such behavior being reported to those in authority. Overall, reciprocal voting detection is a useful starting point for GHHS chapters to mitigate the effect of collusion on the nomination process.

# Figures

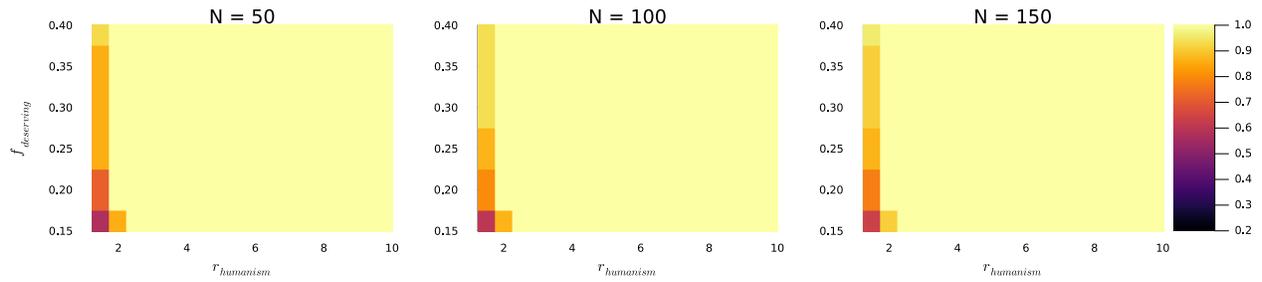

Figure 1: Simulations without collusion. The color is the median fraction of chosen students who were also deserving ($f_{\text{chosen \& deserving}}$) of 100 simulations. This was plotted as a function of class size $N$, fraction of class deserving the award $f_{\text{deserving}}$, and the humanism score $r_{\text{humanism}}$. It is only at the lowest humanism ratios and deserving fraction of the class that there are occasional mistakes in the voting process.

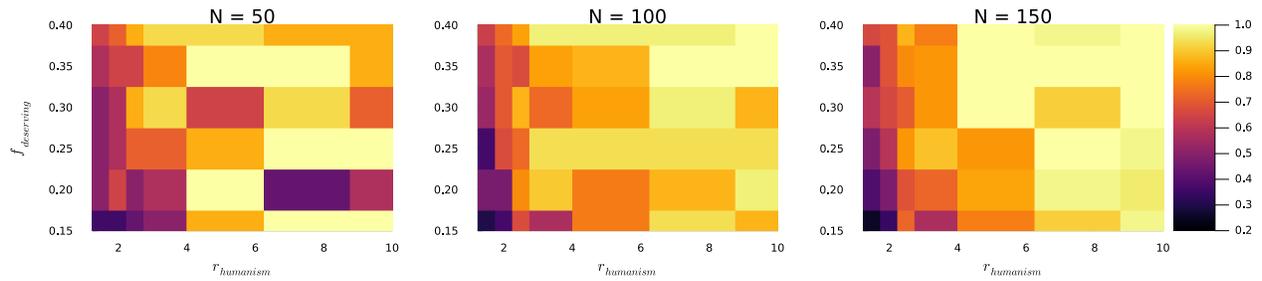

Figure 2: Simulations with collusion. Simulation parameters are the same as in Figure 1, except that there are now friend groups which result in internal voting. The friend groups have sizes which are normally distributed with mean of 10 and standard deviation of 6. There is a marked increase in noise due to collusion among friend groups, resulting in highly-deserving students losing to candidates who are 10 times less deserving (e.g. $r_{\text{humanism}} = 10$). We observe that increasing class size appears to provide some robustness against friend group collusion.

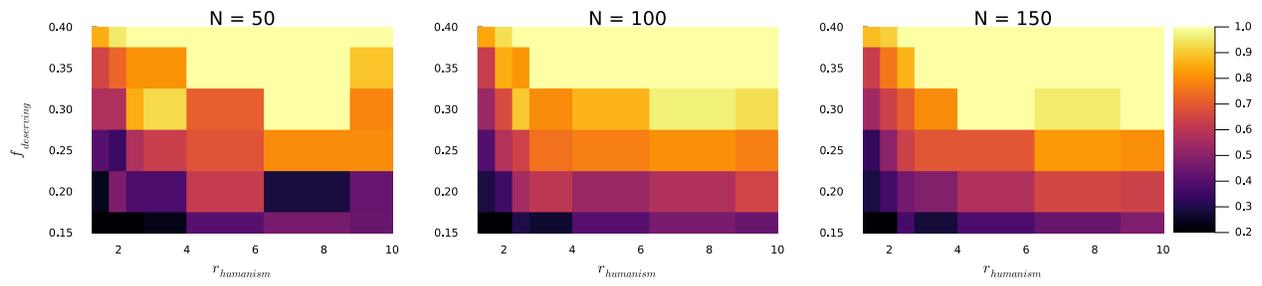

Figure 3: Connected components method of vote adjustment in the presence of collusion. Compared to Figure 2, this method appears to offer benefit at high $f_{\text{deserving}}$, but appears to incorrectly penalize deserving students when $f_{\text{deserving}}$ is low.

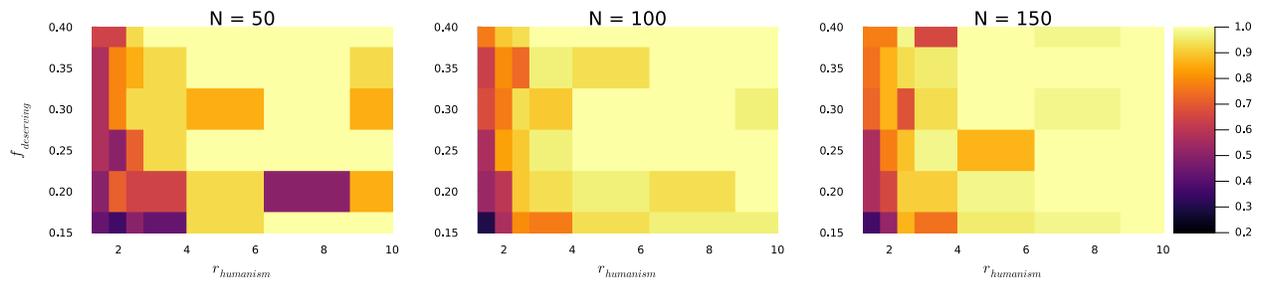

Figure 4: The reciprocal vote discounting method in the presence of collusion. Compared to Figure 2 and 3, this method offers benefit across a wide range of $f_{\text{deserving}}$ and $r_{\text{humanism}}$, which is made more apparent in Figures 5 and 6.

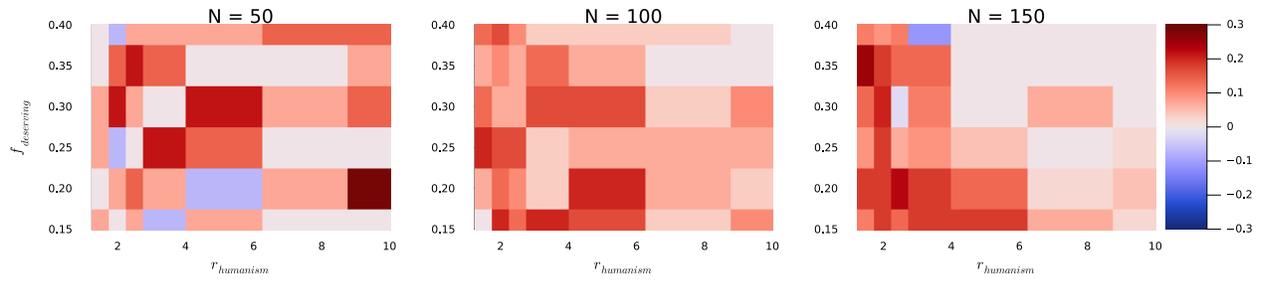

Figure 5: The difference between Figure 4 and Figure 2 (i.e. $f_{\text{chosen \& deserving}}$ for reciprocal vote discounting minus $f_{\text{chosen \& deserving}}$ for simple vote tallying). Reciprocal vote discounting generally results in improved accuracy (more positive change) when identifying deserving students in the presence of collusion.

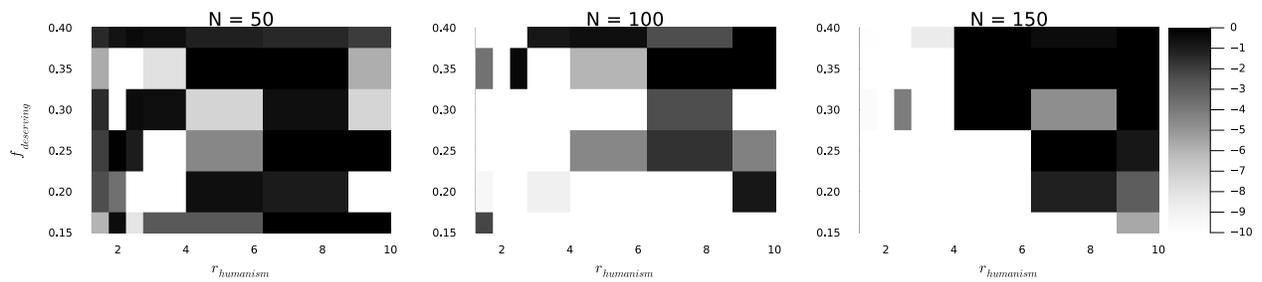

Figure 6: Statistical significance. Colors represent $\log_{10} p$ values of the Mann-Whitney U-test for Figure 5. Since we make 126 comparisons in total, we use the Bonferroni correction so that $\alpha < 0.05$ corresponds to $\log_{10} p < \log_{10} 0.000397 = -3.4$.

**Conflict of interest statement**

The author has no potential conflicts of interest to disclose.

**Code and data availability statement**

All code and data used in this manuscript are included as Supplemental Material.

**Acknowledgments**

This work was not specifically supported by any funding source. The author would like to thank Penn State College of Medicine's Medical Scientist Training Program for its overall support of his career. The views expressed in this article are solely those of the author and do not represent endorsement by any third party.